\documentstyle[aps,preprint]{revtex}
\begin{document}

\draft{}
\title{ Production of pseudovector ($J^{PC}=1^{++}$) heavy quarkonia by virtual
$Z$ boson in $e^+e^-$ collisions
\thanks{This work was supported partly by RFBR-94-02-05 188,
RFBR-96-02-00 548 and INTAS-94-3986}} \author{N.N. Achasov}
\address{Laboratory of Theoretical Physics\\
S.L. Sobolev Institute for Mathematics\\
Universitetskii prospekt, 4\\
Novosibirsk 90,\ \ \  630090,\ \ \ Russia
\thanks{E-mail: achasov@math.nsk.su}} \date{\today}

\maketitle
\begin{abstract}
It is shown that $BR(\chi_{b1}(1P)\rightarrow Z\rightarrow e^+e^-)\simeq
3.3\cdot 10^{-7}$, $BR(\chi_{b1}(2P)\rightarrow Z\rightarrow e^+e^-)\simeq
4.1\cdot 10^{-7}$ and $BR(\chi_{c1}(1P)\rightarrow Z\rightarrow e^+e^-)\simeq
10^{-8}$ that give a good chance to search for the direct production of
pseudovector $^3P_1$ heavy quarkonia in $e^+e^-$ collisions ($e^+e^-\rightarrow
Z\rightarrow ^3P_1$) even at current facilities not to mention $b$ and 
$c-\tau$ factories.
\end{abstract}


   Broadly speaking, production of narrow $p$ wave pseudovector bound states
of heavy quarks $^3P_1$ with $J^{PC}=1^{++}$, like $\chi_{b1}(1p)\,,\,
\chi_{b1}(2p)$ and $\chi_{c1}(1p)$ \cite{pdg-1994}, in $e^+e^-$ collisions via
the virtual intermediate $Z$ boson ($e^+e^-\rightarrow Z\rightarrow ^3P_1$),
could be observed by experiment at least at $b$ and $c-\tau$ factories for
amplitudes of weak interactions grow with energy increase in this energy
regions $\propto G_FE^2$.

   In present paper it is shown that the experimental investigation of this
interesting phenomenon is possible at current facilities.

   First and foremost let us calculate the $^3P_1\rightarrow Z\rightarrow
e^+e^-$ amplitude. We use a handy formalism of description of a
nonrelativistic bound state decays given in the review \cite{novikov-1978}.

The Feynman amplitude describing a free quark-antiquark annihilation into
the $e^+e^-$ pair has the form
\begin{eqnarray}
&& M(\bar Q(p_{\bar Q})Q(p_Q)\rightarrow Z\rightarrow e^+(p_+)e^-(p_-))=
\frac{\alpha\pi}{2\cos^2\theta_W\sin^2\theta_W}\frac{1}{E^2-m^2_Z}
j^{e\,\alpha}j^Q_\alpha= \nonumber\\
&&\frac{\alpha\pi}{2\cos^2\theta_W\sin^2\theta_W}\frac{1}{E^2-m^2_Z}
\bar e(p_-)[(-1+4\sin^2\theta_W)\gamma^\alpha-\gamma^\alpha\gamma_5]e(-p_+)
\cdot \nonumber\\
&& \bar Q_C(-p_{\bar Q})[(t_3-e_Q\sin^2\theta_W)\gamma_\alpha+
t_3\gamma_\alpha\gamma_5]Q^C(p_Q)
\end{eqnarray} 
where the notation is quite marked unless generally accepted.

One can ignore the vector part of the electroweak electron-positron current $
j^{e\,\alpha}$ in Eq. (1) for $(1-4\sin^2\theta_W)\simeq 0.1$. As for the
electroweak quark-antiquark current $j^Q_\alpha$, only its axial-vector part
contributes to the pseudovector ($1^{++}$) quarkonia annihilation. So, the
amplitude of interest is
\begin{eqnarray}
&& M(\bar Q(p_{\bar Q})Q(p_Q)\rightarrow Z\rightarrow e^+(p_+)e^-(p_-))=
\sigma_Q\frac{\alpha\pi}{4\cos^2\theta_W\sin^2\theta_W}
\frac{1}{m^2_Z}j^{e\,\alpha}_5j^Q_{\alpha\,5}=
\nonumber\\
&& \sigma_Q\frac{\alpha\pi}{4\cos^2\theta_W\sin^2\theta_W}
\frac{1}{m^2_Z}\bar e(p_-)\gamma^\alpha\gamma_5e(-p_+)\bar
Q_C(-p_{\bar Q})\gamma_\alpha\gamma_5Q^C(p_Q)
\end{eqnarray} 
where $\sigma_c=1$ and $\sigma_b=-1$. The term of order of $E^2/m_Z^2$ is
omitted in Eq. (2).

In the c.m. system one can write that
\begin{eqnarray}
&& M(\bar Q(p_{\bar Q})Q(p_Q)\rightarrow Z\rightarrow e^+(p_+)e^-(p_-))\simeq
-\sigma_Q\frac{\alpha\pi}{4\cos^2\theta_W\sin^2\theta_W}\frac{1}{m^2_Z}
j^e_{i\,5}\,j^Q_{i\,5}=\nonumber\\
&& -\sigma_Q\frac{\alpha\pi}{4\cos^2\theta_W\sin^2\theta_W}
\frac{1}{m^2_Z}\bar e(p_-)\gamma_i\gamma_5e(-p_+)\bar Q_C(-p_{\bar Q})
\gamma_i\gamma_5Q^C(p_Q)=M({\bf p})
\end{eqnarray} 
ignoring the term of order of $2m_e/E$ ($j^e_{0\,5}=(2m_e/E)j_5$ in the c.m.
system). Hereafter the three-momentum ${\bf p}={\bf p_Q}=-{\bf p_{\bar Q}}$
in the c.m. system.

To construct the effective Hamiltonian for the $1^{++}$ quarkonium
annihilation into the $e^+e^-$ pair one expresses the axial-vector
quark-antiquark current $j^Q_{i\,5}$ in Eq. (3) in terms of two-component
spinors of quark $w^\alpha$ and antiquark $v_\beta$ using four-component
Dirac bispinors
\begin{eqnarray}
&& Q^C(p_Q)= Q^CQ(p_Q)=\frac{1}{\sqrt{2m_Q}}\, Q^C\left (\sqrt{\varepsilon
+ m_Q}\,\, w\atop {\sqrt{\varepsilon - m_Q}\,\, ({\bf n}\cdot\mbox{\boldmath
$\sigma$})w}\right)\,,\nonumber\\
&&\bar Q_C(-p_{\bar Q})=Q_C\bar Q(-p_{\bar Q})=-\frac{1}{\sqrt{2m_Q}}\,
Q_C\left(\sqrt{\varepsilon -m_Q}\,\, v({\bf n}\cdot\mbox{\boldmath $\sigma$})
\,,\,\sqrt{\varepsilon +m_Q}\,\, v\right)
\end{eqnarray} 
where $Q^C$ and $Q_C$ are color spinors of quark and antiquark respectively,
${\bf n}={\bf p}/|{\bf p}|$.

As the result
\begin{equation}
j^Q_{i\,5}=\imath\frac{2\sqrt{6}}{2m_Q}\,\varepsilon_{kin}p_k\chi_n \eta_0
\end{equation}
where $\chi_n=v\sigma_nw/\sqrt{2}$ and $\eta_0=Q_CQ^C/\sqrt{3}$.

The spin-factor $\chi_i$ and the color spin-factor $\eta_0$ are contracted as
follows $\chi_i\chi_j=\delta_{ij}$ and $\eta_0\eta_0=1$.

The $^3P_1$ bound state wave function in the coordinate representation has
the form

\begin{equation}
\Psi_j(^3P_1\,,\,\mbox{\boldmath $r$}\,,\,m_A )= \frac{1}{\sqrt{2}}%
\,\eta_0\varepsilon_{jpl}\chi_p\frac{r_l}{r} \sqrt{\frac{3}{4\pi}}\,
R_P(r\,,\,m_A)
\end{equation}
where $r=|{\bf r}|$, $R_P(r,m_A)$ is a radial wave function with the
normalization $\int_0^{\infty}|R_P(r\,,\,m_A)|^2r^2dr=1$, $m_A$ is a mass of
a $^3P_1$ bound state.

For use one needs the $^3P_1$ bound state wave function in the momentum
representation. It has the form
\begin{equation}
\Psi_j(^3P_1\,,\,\mbox{\boldmath $p$}\,,\,m_A )= \frac{1}{\sqrt{2}}
\,\eta_0\varepsilon_{jpl}\chi_p (\psi_P(\mbox{\boldmath $p$}\,,\,m_A ))_l
\end{equation}
where
\begin{equation}
(\psi_P(\mbox{\boldmath $p$}\,,\,m_A ))_l=\sqrt{\frac{3}{4\pi}}\,\int \frac{
r_l}{r}R_P(r\,,\,m_A)\exp\{-\imath (\mbox{\boldmath $p$}\cdot
\mbox{\boldmath $r$)}\}d^3r\,.
\end{equation}

The amplitude of the $^3P_1$ bound state $\rightarrow e^+e^-$ annihilation
is given by
\begin{equation}
M(A_j\rightarrow e^+e^-)\equiv\int M(\mbox{\boldmath $p$}) \Psi_j(^3P_1\,,\,%
\mbox{\boldmath $p$}\,,\,m_A)\frac{d^3p}{(2\pi)^3}
\end{equation}
where $A_j$ stands for a $^3P_1$ state.

As seen from Eq. (3) to find the $M(A_j\rightarrow e^+e^-)$ amplitude one
needs to calculate the convolution and contraction of a quark-antiquark pair
axial-vector current with a $^3P_1$ bound state wave function
\begin{eqnarray}
&& \int j^Q_{i\,5}\Psi_j(^3P_1\,,\,\mbox{\boldmath $p$}\,,\,m_A)
\frac{d^3p}{(2\pi)^3}=\imath\frac{2\sqrt{6}}{m_A}\,\varepsilon_{kin}
\chi_n\eta_0\int p_k\cdot\Psi_j(^3P_1\,,\,\mbox{\boldmath $p$}\,,\,m_A)
\frac{d^3p}{(2\pi)^3}=\nonumber\\
&& \imath\delta_{ij}\frac{4}{\sqrt{3}}\,\frac{1}{m_A}
\int p_k(\psi_P(\mbox{\boldmath $p$}\,,\,m_A))_k\frac{d^3p}{(2\pi)^3}=
\delta_{ij}2\frac{3}{\sqrt{\pi}}\,\frac{1}{m_A}R_P^\prime(0\,,\,m_A)
\end{eqnarray}  
where $R^\prime_P(0\,,\,m_A)=dR_P(r\,,\,m_A)/dr|_{r=0}$. Deriving Eq. (10)
we put, as it usually is, $2m_Q=m_A$ and took into account that $
R_P(r\,,\,m_A) \rightarrow rR^\prime_P(0\,,\,m_A)$ when $r\rightarrow 0$.

So,
\begin{equation}
M(A_j\rightarrow e^+e^-)=-\sigma_Q3\sqrt{\pi}\,\frac{\alpha}{2\cos^2\theta_W
\sin^2\theta_W}\frac{1}{m^2_Z}\frac{1}{m_A}R^\prime_P(0) \bar
e(p_-)\gamma_j\gamma_5e(-p_+)\,.
\end{equation}

The width of the $A\rightarrow e^+e^-$ decay
\begin{eqnarray}
&&\Gamma (A\rightarrow e^+e^-)=\frac{1}{3}\sum_{j\,e^+e^-}\int |M(A_j
\rightarrow e^+e^-)|^2(2\pi)^4\delta^4(m_A-p_--p_+)\frac{d^3p_+}{(2\pi)^3}
\frac{d^3p_-}{(2\pi)^3}\simeq\nonumber\\
&&\frac{1}{3}\sum_{j\,e^+e^-}\int |M(A_j\rightarrow e^+e^-)|^2
\frac{1}{8\pi}\simeq\nonumber\\
&&\frac{\alpha^2}{32}\frac{3}{\cos^4\theta_W\sin^4\theta_W}
\frac{1}{m_Z^4}\frac{1}{m^2_A}|R^\prime_P(0\,,\,m_A)|^2Sp (\hat p_+\gamma_j
\gamma_5\hat p_-\gamma_j\gamma_5)\simeq\nonumber\\
&&\frac{\alpha^2}{8}\frac{1}{\cos^4\theta_W\sin^4\theta_W}\frac{1}{m_Z^4}
|R^\prime_P(0\,,\,m_A)|^2\simeq 12.3\alpha^2\frac{1}{m_Z^4}|R^\prime_P(0\,,
\,m_A)|^2
\end{eqnarray} 
where terms of order of $(2m_e/m_A)^2$ are omitted, $\sin^2\theta_W=0.225$
is put, the normalization $\bar e(p_-)e(p_-)=2m_e$ and $\bar
e(-p_+)e(-p_+)=-2m_e$ is used. Note that for the quark and antiquark we use
the normalization $\bar Q(p_Q)Q(p_Q)=1$ and $\bar Q(-p_{\bar Q})Q(-p_{\bar
Q})=-1$, see Eq. (4).

   To estimate a possibility of the $\chi_{c1}(1P)$, $\chi_{b1}(1P)$ and
$\chi_{b1}(2P)$ production in $e^+e^-$ collisions it needs to estimate the
branching ratio $BR(A\rightarrow e^+e^-)$.

   In a logarithmic approximation \cite{novikov-1978,barbieri-1976} the decay
of the $^3P_1$ level into hadrons is caused by the decays $^3P_1\rightarrow
g + q\bar q$ where $g$ is gluon and $q\bar q$ is a pair of light quarks:
$u\bar u,\, d\bar d,\,s\bar s$ for $\chi_{c1}(1P)$ and $u\bar u,\,d\bar
d,\,s\bar s,\, c\bar c$ for $\chi_{b1}(1P)$ and $\chi_{b1}(2P)$ . The
relevant width \cite{novikov-1978,barbieri-1976}
\begin{equation}
\Gamma_{log}\left(^3P_1\equiv A\rightarrow gq\bar q\right)\simeq \frac{N}{3}
\frac{128}{3\pi}\frac{\alpha^3_s}{m_A^4}|R^\prime_P(0\,,\,m_A)|^2
\ln\frac{m_AR(m_A)}{2}
\end{equation}
  where N is the number of the light quark flavors and $R(m_A)$ is the
quarkonium radius. Using Eqs. (12) and (13) one gets that
\begin{equation}
BR(A\rightarrow e^+e^-)\simeq \frac{3}{N}0.9\frac{\alpha^2}{\alpha_s^3}
\left (\frac{m_A}{m_Z}\right)^4\frac{1}{\ln\left(m_AR(m_A)/2\right)}\left[1-
\sum_VBR(A\rightarrow\gamma V)\right]
\end{equation}
where the radiative decays $\chi_{c1}(1P)\rightarrow\gamma J/\psi$,
$\chi_{b1}(1P)\rightarrow\gamma \Upsilon (1S)$,
$\chi_{b1}(2P)\rightarrow\gamma \Upsilon (1S)$ and
$\chi_{b1}(2P)\rightarrow\gamma \Upsilon (2S)$
are taken into account.

A convention \cite{novikov-1978} uses that $L(m_A)=
\ln\left(m_AR(m_A)/2\right)\simeq 1$ for $\chi_{c1}(1P)$, i.e. when $m_A=
3.51\,GeV$ \cite{pdg-1994}. As for $\chi_{b1}(1P)$, $m_A=9.89\,GeV$ \cite
{pdg-1994}, and $\chi_{b1}(2P)$, $m_A=10.2552\,GeV$ \cite{pdg-1994}, it
depends on the $m_A$ behavior of the quarkonium radius $R(m_A)$. For example,
the coulomb-like potential gives that $R(m_A)\sim 1/m_A$ and the logarithm
practically does not increase, $L(3.51\,GeV)\simeq L(9.89\,GeV)\simeq
L(10.2552\,GeV)\simeq 1$. Alternatively, the harmonic oscillator potential
gives $R(m_A)\sim 1/\sqrt{m_A\omega_0}$, where $\omega_0\simeq 0.3\,GeV$, that
leads to $L(9.89\,GeV)\simeq L(10.2552\,GeV)\simeq 1.5$. To be conservative
one takes $L(9.89\,GeV)=L(10.2552\,GeV)=2$.

So, putting $\alpha_s(3.51\,GeV)=0.2\,,\,BR(\chi_{c1}(1P)\rightarrow\gamma
J/\psi)=0.27$ \cite{pdg-1994}, $N=3\,,\,L(3.51\,GeV)=1\,,\,m_A=
m_{\chi_{c1}(1P)}=3.51\,GeV$ and $m_Z=91.2\,GeV$ one gets from Eq. (14) that
\begin{equation}
BR(\chi_{c1}(1P)\rightarrow e^+e^-)=0.96\cdot 10^{-8}\,.
\end{equation}

   Putting $\alpha_s(9.89\,GeV)=\alpha_s(10.2552\,GeV)=0.17\,,\,
BR(\chi_{b1}(1P)\rightarrow\gamma \Upsilon (1S))=0.35 \cite{pdg-1994}\,,
BR(\chi_{b1}(2P)\rightarrow\gamma \Upsilon (1S))+
BR(\chi_{b1}(2P)\rightarrow\gamma \Upsilon (2S))=0,085+0.21=0.295
\cite{pdg-1994}\,,\, N=4\,,\,L(9.89\,GeV)=L(10.2552\,GeV) =2\,,\,m_A=
m_{\chi_{b1}(1P)}=9.89\,GeV\,,\,m_A=m_{\chi_{b1}(2P)}=10.2552\,GeV$ and
$m_Z=91.2\,GeV$ one gets from Eq. (14) that
\begin{eqnarray}
&&BR(\chi_{b1}(1P)\rightarrow e^+e^-)=3.3\cdot 10^{-7}\,,\\ \nonumber
&&BR(\chi_{b1}(2P)\rightarrow e^+e^-)=4.1\cdot 10^{-7}\,.
\end{eqnarray}

  Let us discuss  possibilities to measure the branching ratios under
consideration .

The cross section of a reaction $e^+e^-\rightarrow A\rightarrow out$ at
resonance peak \cite{pdg-1994}
\begin{equation}
\sigma (A)\simeq 1.46\cdot 10^{-26}BR(A\rightarrow e^+e^-)BR(A\rightarrow
out) \left(\frac{GeV}{m_A}\right)^2\,cm^2\,.
\end{equation}

So, for the $\chi_{c1}(1P)$ state production
\begin{equation}
\sigma \left(\chi_{c1}(1P)\right)\simeq 1.14\cdot 10^{-35}BR\left
(\chi_{c1}(1P) \rightarrow out\right)\,cm^2
\end{equation}
and for the production of the $\chi_{b1}(1P)$ and $\chi_{b1}(2P)$ states
\begin{eqnarray}
&&\sigma \left(\chi_{b1}(1P)\right)\simeq 4.8\cdot
10^{-35}BR\left(\chi_{b1}(1P) \rightarrow out\right)\,cm^2\,,\\ \nonumber
&&\sigma \left(\chi_{b1}(2P)\right)\simeq 5.6\cdot
10^{-35}BR\left(\chi_{b1}(2P) \rightarrow out\right)\,cm^2\,.
\end{eqnarray}

In general, the visible cross section at the peak of the narrow resonances
like $J/\psi\,,\,\Upsilon (1S)$ and so on is suppressed by a factor of order
of $\Gamma_{tot}/\Delta E$ where $\Delta E$ is an energy spread. But,
fortunately, the $\chi_{c1}(1P)$ resonance width equal to 0.88 $MeV$ \cite
{pdg-1994} is not small in comparison with energy spreads of current
facilities, for example, $\Delta E\simeq 2\,MeV$ at BEPC (China), see \cite
{pdg-1994}. Taking into account that the luminosity at BEPC \cite{pdg-1994} is
equal to $10^{31}\,cm^{-2} s^{-1}\,,\,\Gamma_{tot}(\chi_{c1}(1P))/\Delta E
\simeq 0.44$ and the cross section of the $\chi_{c1}(1P)$ production is equal
to $1.14\cdot 10^{-35}\,cm^2$, see Eq. (18), one can during an effective year
($10^7$ seconds) working produce 501 $\chi_{c1}(1P)$ states.

Note that such a number of $\chi_{c1}(1P)$ states gives 135 (27\%) unique
decays $\chi_{c1}(1P)\rightarrow\gamma J/\psi$.

The $c-\tau $ factories (luminosity $\sim 10^{33}\,cm^{-2}s^{-1}$) could
produce several tens of thousands of the $\chi _{c1}(1P)$ states.

As for the $\chi_{b1}(1P)$ state, its width is unknown up to now \cite
{pdg-1994}. Let us estimate it using the $\chi_{c1}(1P)\,,\,J/\psi\,,\,
\Upsilon(1S)$ widths, and the quark model.

   In the quark model
\begin{equation}
\Gamma\left(^3S_1\equiv V\rightarrow ggg\right)=\frac{40}{81\pi}\left(\pi^2-9
\right)\frac{\alpha^3_s}{m_V^2}|R_S(0\,,\,m_V)|^2
\end{equation}

One gets from Eqs. (13) and (20) that
\begin{eqnarray}
&&\frac{\Gamma (A\rightarrow gq\bar q)}{\Gamma (V\rightarrow ggg)}=
\nonumber\\[3pt]
&&\frac{\Gamma_{tot}(A)}{\Gamma_{tot}(V)}\frac{BR(A\rightarrow hadrons)}{
[BR(V\rightarrow hadrons)- BR(V\rightarrow virtual\,\gamma\rightarrow
hadrons)]}=\nonumber\\[3pt]
&& 99.4\frac{N}{3}\left(\frac{m_V}{m_A}\right)^2
\left|\frac{R^{\prime}_P(0\,,\,m_A)}{m_AR_S(0\,,\,m_V)}\right|^2
\ln\frac{m_AR(m_A)}{2}\,.
\end{eqnarray} 

So,
\begin{eqnarray}
&&\frac{\Gamma_{tot}(\chi_{b1}(1P))}{\Gamma_{tot}(\Upsilon (1S))}=
0.53\frac{\Gamma_{tot}(\chi_{c1}(1P))}{\Gamma_{tot}(J/\psi)}\left|\frac{
R^\prime_P(0\,,\,m_{\chi_{b1}(1P)})R_S(0\,,\,m_{J/\psi})}{R^\prime_P(0\,,
\,m_{\chi_{c1}(1P)})R_S(0\,,\,m_{\Upsilon (1S)})}\right|^2=\nonumber\\[6pt]
&&5.3\left|\frac{R^\prime_P(0\,,\,m_{\chi_{b1}(1P)})R_S(0\,,\,m_{J/\psi})}
{R^\prime_P(0\,,\,m_{\chi_{c1}(1P)})R_S(0\,,\,m_{\Upsilon (1S)})}\right|^2\,.
\end{eqnarray} 
Calculating Eq. (22) one used data from \cite{pdg-1994},
$BR(\Upsilon (1S)\rightarrow hadrons)- BR(\Upsilon (1S)\rightarrow virtual\,
\gamma\rightarrow hadrons)=0.83$, $BR(J/\psi\rightarrow hadrons)-
BR(J/\psi\rightarrow virtual\, \gamma\rightarrow hadrons)=0.69\,,\,
\Gamma_{tot}(\chi_{c1}(1P))/\Gamma_{tot}(J/\psi)=10$, and $L(m_{b1(1P)})
/L(m_{c1(1P)})=2$ as in the foregoing.

The unknown factor in Eq. (22) depends on a model. In the Coulomb-like
potential model it is
\begin{equation}
\left|\frac{R^\prime_P(0\,,\,m_{\chi_{b1}(1P)})R_S(0\,,\,m_{J/\psi})}
{R^\prime_P(0\,,\,m_{\chi_{c1}(1P)})R_S(0\,,\,m_{\Upsilon (1S)})}\right|^2=
\left(\frac{m_{\chi_{b1}(1P)}}{m_{\chi_{c1}(1P)}}\right)^5 \left(\frac{%
m_{J/\psi}}{m_{\Upsilon (1S)}}\right)^3=6.2 \,.
\end{equation}

In the harmonic oscillator potential it is
\begin{equation}
\left|\frac{R^\prime_P(0\,,\,m_{\chi_{b1}(1P)})R_S(0\,,\,m_{J/\psi})}
{R^\prime_P(0\,,\,m_{\chi_{c1}(1P)})R_S(0\,,\,m_{\Upsilon (1S)})}\right|^2=
\left(\frac{m_{\chi_{b1}(1P)}}{m_{\chi_{c1}(1P)}}\right)^{2.5} \left(\frac{%
m_{J/\psi}}{m_{\Upsilon (1S)}}\right)^{1.5}=2.5 \,.
\end{equation}

To be conservative one takes Eq. (24). Thus one expects
\begin{equation}
\Gamma_{tot}(\chi_{b1}(1P))\simeq 13\Gamma_{tot}(\Upsilon (1S))\simeq 0.695\,
MeV\,.
\end{equation}

Let us estimate a number of the $\chi_{b1}(1P)$ states which can be produced
at CESR (Cornell) \cite{pdg-1994}. Taking into account that luminosity at
CESR is equal to $10^{32}\,cm^{-2}s^{-1}\,,\,\Delta E\simeq 6\,MeV\,,\,
\Gamma_{tot}(\chi_{b1}(1P))/\Delta E\simeq 0.12$ and the cross section of the
$\chi_{b1}(1P)$ production is equal to $4.8\cdot 10^{-35}\,cm^2$, see Eq.
(19), one can during an effective year ( $10^7$ seconds) working produce 5622
$\chi_{b1}(1P)$ states. This number of the $\chi_{b1}(1P)$ states gives 1968
(35\%) unique decays $\chi_{b1}(1P)\rightarrow\gamma\Upsilon (1S)$.

At VEPP-4M (Novosibirsk) \cite{pdg-1994} one can produce a few hundreds of
the $\chi_{b1}(1P)$ states.

As for the $b$ factories with luminosities $10^{33}\,cm^{-2}s^{-1}$ and
$10^{34}\,cm^{-2}s^{-1}$ \cite{pdg-1994}, they could produce tens and
hundreds of thousands of the $\chi_{b1}(1P)$ states.

 As for the $\chi_{b1}(2P)$ state, it is impossible to estimate its width by
the considered way. The point is that the $\chi_{c1}(2P)$ is unknown up to now
\cite{pdg-1994} (probably, this state lies above the threshold production of
the $D\bar D^* + D^*\bar D$ heavy quarkonia). Nevertheless, it seems 
reasonable that $\Gamma_{tot}(\chi_{b1}(2P))\sim 1\,MeV$ as in the case of the
$\chi_{b1}(1P)$ state. That is why it is reasonable to believe, see Eq. (19), 
that there is a good chance to search for the direct production of the 
$\chi_{b1}(2P)$ state as in the case of the $\chi_{b1}(1P)$ state.

So, the current facilities give some chance to observe the $\chi_{c1}(1P)$
state production in the $e^+e^-$ collisions and to study the production of
the $\chi_{b1}(1P)$ and $\chi_{b1}(2P)$ states in the $e^+e^-$ collisions in
sufficient detail.

The $c-\tau$ and $b$ factories would give possibilities to study in the
$e^+e^-$ collisions the $\chi_{c1}(1P)$ state production in sufficient detail
and the production of the $\chi_{b1}(1P)$ and $\chi_{b1}(2P)$ states in depth.
Probably, it is possible to observe the $\chi_{c1}(2P)$ state production at
the $c-\tau$ factories.

The fine effects considered above are essential not only to the
understanding of the quark model but can be used for identification of the
$\chi_{b1}(1P)$ and $\chi_{b1}(2P)$  states because the angular momentum $J$
of the states named as $\chi_{b1}(1P)$ and $\chi_{b1}(2P)$ needs confirmation
\cite{pdg-1994}.

I would like to thank V.V. Gubin, A.A. Kozhevnikov and G.N. Shestakov for
discussions.

\end{document}